\documentclass[12pt,letterpaper]{extarticle}
\usepackage{ae,aecompl}
\usepackage{avant}

\usepackage[T1]{fontenc}
\usepackage[latin9]{inputenc}
\usepackage{enumitem}
\usepackage{amsmath}
\usepackage{amsthm}
\usepackage{amssymb}
\usepackage{setspace}
\usepackage[authoryear]{natbib}
\makeatletter

\theoremstyle{definition}
\newtheorem{defn}{\protect\definitionname}
\theoremstyle{plain}
\newtheorem{thm}{\protect\theoremname}
\theoremstyle{plain}
\newtheorem{prop}{\protect\propositionname}
\theoremstyle{plain}
\newtheorem{lem}{\protect\lemmaname}
\theoremstyle{remark}
\newtheorem{claim}{\protect\claimname}

\usepackage[bottom]{footmisc}
\usepackage{bm}
\usepackage{pdfsync}
\providecommand{\definitionname}{Definition}
\providecommand{\lemmaname}{Lemma}
\providecommand{\theoremname}{Theorem}

\makeatother

\providecommand{\claimname}{Claim}
\providecommand{\definitionname}{Definition}
\providecommand{\lemmaname}{Lemma}
\providecommand{\propositionname}{Proposition}
\providecommand{\theoremname}{Theorem}

\begin{document}
\title{(Non-)Obvious Manipulability of Rank-Minimizing Mechanisms}
\author{Peter Troyan \\
Department of Economics \\
University of Virginia\thanks{Address: P.O. Box 400182, Charlottesville, VA, 22904. Email: troyan@virginia.edu.
I would like to thank Josue Ortega for comments on this paper, as
well the Associate Editor and two anonymous referees, whose comments
have greatly improved it.}\\
}
\maketitle
\begin{abstract}
In assignment problems, the rank distribution of assigned objects
is often used to evaluate match quality. Rank-minimizing (RM) mechanisms
directly optimize for average rank. While appealing, a drawback is
RM mechanisms are not strategyproof. This paper investigates whether
RM satisfies the weaker incentive notion of non-obvious manipulability
\citep[NOM,][]{troyan2020obvious}. I show any RM mechanism with full
support---placing positive probability on all rank-minimizing allocations---is
NOM. In particular, uniform randomization satisfies this condition.
Without full support, whether an RM mechanism is NOM or not depends
on the details of the selection rule.

\pagebreak{}
\end{abstract}

\section{Introduction}

Many institutions make assignments by collecting agents' ordinal preferences
over the possible alternatives and using them as an input to a rule
that outputs an assignment. Common examples include public school
choice \citep{abdul2003school}, medical residency matching \citep{roth/peranson:99},
teacher assignment \citep{combe2018design}, course allocation \citep{budish/cantillon:10,budish2014changing},
and refugee resettlement \citep{delacretaz2016refugee}. A natural
metric to measure the success of the outcome is the rank distribution:
how many students get their first choice, how many get their second
choice, and so on. Indeed, many school districts such as those in
New York City and San Francisco publicly release statistics on the
ranks as a measure of the goodness of the match.

In the context of school choice specifically, most cities rely on
either some version of Gale and Shapley's celebrated deferred acceptance
mechanism \citep{Gale:AMM:1962} or on the so-called Boston mechanism
(also sometimes called the immediate acceptance mechanism) to determine
the assignment, and then evaluate the rank distribution produced by
these mechanisms.\footnote{There is another class of mechanisms based on the top trading cycles
(TTC) algorithm of \citet{shapley/scarf:74} that has been studied
extensively in the theoretical literature, but has found few adherents
in practice. The only use of it in a real-world school choice setting
that I am aware of is New Orleans, where it was used for one year
before being abandoned \citep{Abdulkadiro_lu_2020}.} However, given the importance placed on the rank distribution, it
is also natural to consider mechanisms that optimize directly for
this objective. Indeed, \citet{featherstone2014rank} notes that Teach
for America does exactly this, and uses the rank distribution when
\emph{selecting }its assignment. \citet{featherstone2014rank} is
also one of the few papers that has undertaken a serious analysis
of mechanisms based explicitly on the rank distribution (a few other
papers in this relatively small but growing literature are discussed
below).

While using mechanisms that select assignments based explicitly on
the rank distribution is naturally appealing, an important consideration
in any mechanism design problem is the incentives of the agents. Indeed,
one of the most appealing properties arguing for the use of DA-based
mechanisms is that they generally give agents strong incentives to
report their true preferences. On the other hand, Proposition 10 of
\citet{featherstone2014rank} shows that no ordinal assignment mechanism
is both rank-efficient (a refinement of Pareto efficiency that he
defines that takes into account the rank distribution) and strategyproof.

However, strategyproofness is a very demanding property, and just
because a mechanism \emph{can }be manipulated does not mean that it
is \emph{will} be manipulated. As part of a recent strand of literature
on ``obviousness'' in mechanism design, \citet{troyan2020obvious}
introduce the concept of \emph{non-obvious manipulability (NOM) }as
a way to relax strategyproofness.\emph{ }They use their definition
to taxonomize non-strategyproof mechanisms into two classes: those
that are obviously manipulable (such as the Boston mechanism and pay-as-bid
auctions) and thus are likely to be easily manipulated in practice,
and those that are non-obviously manipulable (such as school-proposing
DA, Kesten's (\citeyear{kesten:QJE:2010}) efficiency-adjusted DA,
and uniform price auctions) which, while formally manipulable, have
manipulations that are more difficult for cognitively-limited agents
to recognize and enact successfully.\footnote{Experimentally, \citet{cerrone2022school} find high rates of truth-telling
in the non-strategyproof, but also not obviously manipulable, EADA
mechanism. }

In this paper, I consider a canonical assignment model in which there
is a set of agents (such as students) to be assigned to a set of objects
(such as schools), each of which has some fixed capacity. Agents participate
in a mechanism in which they report their preferences over the objects.
I consider the class of \emph{rank-minimizing (RM) mechanisms, }which
take the reported agent preferences, and implement an assignment that
minimizes the average rank of the objects received. As there can in
general be many allocations that minimize the average rank at a given
preference profile, there are many possible RM mechanisms, depending
on how these ties are broken. Because the aforementioned result of
\citet{featherstone2014rank} implies that no RM mechanism is strategyproof,
I consider instead the weaker notion of NOM from \citet{troyan2020obvious},
and ask whether RM mechanisms satisfy this property.

An issue that arises in answering this question is that Troyan and
Morrill's definition applies only to deterministic mechanisms that
select a single outcome at each preference profile. Because there
may be many allocations that minimize the average rank for a given
preference profile, it is more natural to model a mechanism as returning
a probability distribution over such allocations, and use an extension
of NOM for probabilistic mechanisms that was first proposed by \citet{demeulemeester2022rawlsian}. 

My main result (Theorem \ref{thm:RM-is-NOM}) shows that any\emph{
}RM mechanism with full support is not obviously manipulable. A full
support RM mechanism is one that places strictly positive probability
weight on every allocation that is rank-minimizing at the submitted
preference profile. For instance, one natural way to run an RM mechanism
is to select uniformly at random from the set of all RM allocations,
which we call the \emph{uniform rank-minimizing (URM) mechanism}.\footnote{This is a commonly used implementation of RM mechanisms in the literature;
see, e.g., \citet{Nikzad2022Rank} or \citet{Ortega2022ImprovingEfficiency}.} This is a full support mechanism, and so Theorem \ref{thm:RM-is-NOM}
implies that the URM mechanism is NOM. 

After showing Theorem \ref{thm:RM-is-NOM}, I investigate whether
NOM extends beyond the full support assumption. This may be important,
for instance, to a designer who may not view all RM allocations as
equal, but rather wants to optimize some further objective within
this class. For instance, if a designer wants to advantage agents
who belong to certain groups within the class of RM allocations, she
might want to place zero probability on RM allocations where these
agents do worse, and higher probability on RM allocations where they
do better. 

Without full support, whether an RM mechanism is NOM will depend on
the details of the selection rule. Consider the following mechanism,
called the \emph{rank-minimizing serial dictatorship (RM-SD)}: Fix
an exogenous ordering of the agents, and ask the agents in this order
to choose their favorite (remaining) object that they could be assigned
at any RM allocation that is also consistent with the choices of the
earlier agents. This is a deterministic mechanism (it always produces
a single deterministic allocation), and so does not have full support.
I show that if each object has capacity 1 (what I call \emph{unit
capacity markets}), the RM-SD mechanism is NOM (Proposition \ref{prop:RMSD-isNOM}).
However, I also show that if some object has a capacity greater than
1, then the RM-SD mechanism is not NOM (Proposition \ref{prop:RM-SD-is-obvi-manip}).
Further, even in unity capacity markets, there are RM mechanisms without
full support that are not NOM, i.e., that are obviously manipulable
(Proposition \ref{prop:OM-that-is-RM}).\footnote{An important additional distinction between these mechanisms and the
uniform RM mechanism discussed in the previous paragraph is that the
latter treats all of the agents fairly. In particular, URM satisfies
the fairness property of \emph{equal treatment of equals (ETE) }while
the others do not.} 

In sum, the class of RM mechanisms is an appealing class of mechanisms
for policymakers who are interested in the average rank as a desirable
objective. To the extent that the main shortcoming of the RM mechanism
is its lack of strategyproofness, my results suggest that this problem
may be not so severe, so long as the mechanism has full support (e.g.,
the uniform RM mechanism). At the same time, another takeaway from
my results is that designers who desire to use different selection
rules without full support should be prudent in doing so, as this
choice could have consequences for the manipulability of the resulting
mechanism. While more empirical work is needed to test the theory,
given my results and their other appealing properties, RM mechanisms
at the very least seem worthy of further consideration for practical
market design applications.

\subsection*{Related literature }

Besides \citet{featherstone2014rank}, who provides a detailed analysis
of rank efficiency criteria and related mechanisms and was discussed
above, there are only three other papers I am aware of in the economics
literature that analyze RM mechanisms.\footnote{There is a large literature in mathematics and operations research
that has studied related problems, though they are generally not concerned
with incentive issues, which are the focus of this paper. See \citet{krokhmal2009random}
for a survey.} \citet{Nikzad2022Rank} studies large markets, and provides an upper
bound on the expected average rank of rank-minimizing assignments.
\citet{sethuraman2022Rank} provides another proof of the same result.
\citet{Nikzad2022Rank} also shows that the uniform RM mechanism is
Bayesian incentive compatible when agent preference rankings are also
drawn uniformly at random.

Lastly, \citet{Ortega2022ImprovingEfficiency} study the average rank
of RM, DA, and TTC in large markets, both theoretically and using
simulations, and find that RM outperforms DA and TTC on important
dimensions such as efficiency and fairness. They also use data from
secondary school admissions in Hungary to analyze the three mechanisms
in an empirical setting, and find similar support for RM mechanisms.
The data is from DA, a strategyproof mechanism, and they conduct their
analysis of RM mechanisms assuming the DA reports are truthful and
that agents will continue to report these truthful preferences in
a counterfactual in which they play RM, even though it is not strategyproof.
They argue for this approach as follows:
\begin{quote}
In our view, it is unclear whether students would misrepresent their
preferences in RM. The potential gains from manipulation are tiny...and
manipulations are risky and could lead to worse outcomes...Furthermore,
there is evidence of high truth-telling rates in not obviously manipulable
mechanisms \citep{cerrone2022school}. 
\end{quote}
My results can be seen as a formalization of this final point of \citet{Ortega2022ImprovingEfficiency}.

\section{Model}

\label{sec:model}

\subsection{Preferences and allocations}

There is a set of $N$ agents $I=\{i_{1},\ldots,i_{N}\}$ and a set
of $M$ objects $O=\{o_{1},\ldots,o_{M}\}$. Each object $o_{m}$
has a \textbf{capacity }$q_{m}$ which denotes the maximum number
of agents who can be assigned to it. I assume that $\sum_{m=1}^{M}q_{m}\geq N$,
which is common in school choice settings where all students must
be offered a seat at a school, and is without loss of generality if
one of the objects is an ``outside option'' that has enough capacity
for all agents. Each agent has a \textbf{strict preference relation}
$\succ_{i}$ defined on the set of objects $O$, where $o\succ_{i}o'$
denotes that agent $i$ strictly prefers object $o$ to object $o'$.
I use $o\succsim_{i}o'$ when either $o\succ_{i}o'$ or $o=o'$. I
will sometimes refer to $\succ_{i}$ as an agent's \textbf{type}.
I will also write $\succ_{i}:o,o',\ldots,$ to denote an agent who
has preferences such that her favorite object is $o$, second favorite
object is $o'$, and the rest of her preferences can be arbitrary.
The set $\mathcal{P}_{i}$ is agent $i$'s preference domain, which
consists of all strict rankings over $O$, and $\mathcal{P}^{I}=\mathcal{P}_{1}\times\cdots\times\mathcal{P}_{N}$
is the set of all preference profiles for all agents. I write $\succ_{I}=(\succ_{1},\ldots,\succ_{N})\in\mathcal{P}^{I}$
to denote a profile of preferences, one for each agent $i_{1},\ldots,i_{N}$,
and sometimes write $\succ_{I}=(\succ_{i},\succ_{-i})$ to separate
$i$'s preferences $\succ_{i}$ from those of the remaining agents,
$\succ_{-i}$. 

For an agent $i$ with type $\succ_{i}$, I define $r_{i}(o)=|\{o'\in O:o'\succsim_{i}o\}|$
to be the \textbf{rank }of object $o$ according to $i$'s preferences;
in other words, $i$'s favorite object has rank 1, $i$'s second-favorite
object has rank 2, etc. While the rank of an object also depends on
an agent's preferences $\succ_{i}$, for readability, I suppress this
from the notation. Also, notice that I use the convention that lower
numbers correspond to more preferred objects.

A (deterministic) \textbf{allocation }$\alpha:I\rightarrow O$ is
a function that assigns each agent to an object. I use $\alpha_{i}$
to denote the object assigned to agent $i$ in allocation $\alpha$.
Any allocation must satisfy $|\{i\in I:\alpha_{i}=o\}|\leq q_{m}$
for all $o_{m}\in O$, i.e., each object cannot be assigned to more
agents than its capacity. Let $\mathcal{A}$ be the set of all possible
allocations. A \textbf{random allocation }$\mu:\mathcal{A}\rightarrow[0,1]$
is a probability distribution over $\mathcal{A}$, where $\sum_{\alpha\in\mathcal{A}}\mu(\alpha)=1$.
It is necessary to introduce random allocations to be able to deal
with tie-breaking when there are many possible deterministic allocations
that minimize the average rank. Let $\mathcal{M}$ be the set of random
allocations.

Given a preference profile $\succ_{I}$, denote the \textbf{average
rank }of any (deterministic) allocation $\alpha$ (with respect to
$\succ_{I}$) by\footnote{Once again, $\bar{r}(\alpha)$ depends on the preference profile $\succ_{I}$,
but this is suppressed to avoid notational clutter.} 
\[
\bar{r}(\alpha)=\frac{1}{N}\sum_{i\in I}r_{i}(\alpha_{i}),
\]
and let 
\[
\bar{\mathcal{A}}(\succ_{I})=\{\alpha\in\mathcal{A}:\bar{r}(\alpha)\leq\bar{r}(\alpha')\text{ for all }\alpha'\in\mathcal{A}\}.
\]
We call $\bar{\mathcal{A}}(\succ_{I})$ the set of \textbf{rank-minimizing
deterministic allocations} (with respect to $\succ_{I}$). Since $\mathcal{A}$
is finite, the set $\bar{\mathcal{A}}(\succ_{I})$ is non-empty for
all $\succ_{I}$, though it may contain more than one element. A random
allocation $\mu$ is a \textbf{rank-minimizing random allocation }(with
respect to $\succ_{I})$ if $\alpha\notin\bar{\mathcal{A}}(\succ_{I})$
implies $\mu(\alpha)=0$; in other words, $\mu$ places positive probability
weight on only those deterministic allocations that are rank-minimizing
given the preference profile $\succ_{I}$. Let $\bar{\mathcal{M}}(\succ_{I})$
denote the set of rank-minimizing random allocations. Notice that
a rank-minimizing random allocation may have $\mu(\alpha)=0$ for
some rank-minimizing (deterministic) allocations $\alpha$. When $\mu(\alpha)>0$
for all $\alpha\in\bar{\mathcal{A}}(\succ_{I})$, we call $\mu$ a\textbf{
full-support rank-minimizing random allocation}. This distinction
will be important for the results below. 

\subsection{Mechanisms}

A \textbf{mechanism }is a function $\psi:\mathcal{P}^{I}\rightarrow\mathcal{M}$.
For each preference profile $\succ_{I}\in\mathcal{P}^{I}$, mechanism
$\psi$ returns the random allocation $\psi(\succ_{I})\in\mathcal{M}$.
For any deterministic allocation $\alpha\in\mathcal{A}$, we write
$\psi(\succ_{I})(\alpha)$ to denote the probability that the random
allocation $\psi(\succ_{I})$ places on $\alpha$. A mechanism $\psi$
is a \textbf{rank-minimizing (RM) mechanism }if $\psi(\succ_{I})$
is a rank-minimizing random allocation for all $\succ_{I}$. In other
words, $\psi(\succ_{I})$ places strictly positive probability only
on allocations that are rank-minimizing; formally, $\psi(\succ_{I})\in\bar{\mathcal{M}}(\succ_{I})$
for all $\succ_{I}\in\mathcal{P}^{I}$. If $\psi(\succ_{I})$ is a
further full-support rank-minimizing allocation for all $\succ_{I}$,
we say that $\psi$ is a \textbf{full-support RM mechanism}. One natural
tie-breaking rule is to choose uniformly at random from the entire
set $\bar{\mathcal{A}}(\succ_{I})$ of rank-minimizing deterministic
allocations. This results in a particular full-support RM mechanism
that we call the \textbf{uniform rank-minimizing (URM) mechanism}.
Of course, other tie-breaking rules---both with full support and
without---are possible as well, and we discuss some of these below.

\subsection{Non-obvious manipulability}

\citet{troyan2020obvious} define non-obvious manipulability for deterministic
direct revelation mechanisms. Informally, a mechanism is not obviously
manipulable if, for every agent and every type $\succ_{i}$, and every
possible misreport $\succ_{i}'\neq\succ_{i}$: (i) the worst-case
outcome under $\succ_{i}$ is weakly better than the worst-case outcome
under $\succ_{i}'$ and (ii) the best-case outcome under $\succ_{i}$
is weakly better than the best-case outcome under $\succ_{i}'$, where
the worst and best cases are taken over all possible reports of the
other agents, $\succ_{-i}$. They provide a characterization of non-obvious
manipulations as those that cannot be recognized by a cognitively
limited agent, and classify mechanisms as either obviously manipulable
or not obviously manipulable in a wide variety of settings that is
in line with empirical evidence.

Formally, \citet{troyan2020obvious}'s definition applies only to
deterministic mechanisms, and so it must be extended to deal with
the probabilistic mechanisms used in this paper. The natural extension
is to simply treat Nature as another player, and calculate the worst
and best possible outcomes over both all possible reports of the other
agents $\succ_{-i}$ as well as all possible realizations of random
draws by Nature.\footnote{Non-obvious manipulability shares a similar motivation with the seminal
paper of \citet{Li-OSP} on obvious dominance (and indeed, the model
of cognitive limitations used by \citet{troyan2020obvious} to characterize
obvious manipulations is the same used by \citet{Li-OSP} to characterize
obvious dominance). In his paper, Li writes: ``Weak dominance treats
chance moves and other players asymmetrically...By contrast, obvious
dominance treats chance moves and other players symmetrically''. In
the original model of \citet{troyan2020obvious}, Nature (or, what
\citet{Li-OSP} refers to as ``chance'') does not play a role, but
here, I must incorporate it. The definition I use continues in the
same spirit of \citet{Li-OSP}, by treating Nature and other players
symmetrically.} This extension of \citet{troyan2020obvious} to random mechanisms
first appears (to my knowledge) in \citet{demeulemeester2022rawlsian},
and the definition below is equivalent to theirs. When the mechanism
itself is deterministic, it reduces to the definition of \citet{troyan2020obvious}. 

Formally, given a random allocation $\mu\in\mathcal{M}$, let $\text{supp}(\mu)=\{\alpha\in\mathcal{A}:\mu(\alpha)>0\}$
be the \textbf{support }of $\mu$ and define: 
\[
\bar{\rho}_{i}(\mu)=\max_{\alpha\in\text{supp}(\mu)}r_{i}(\alpha_{i})
\]

\[
\text{\ensuremath{\underbar{\ensuremath{\rho}}}}_{i}(\mu)=\min_{\alpha\in\text{supp}(\mu)}r_{i}(\alpha_{i}).
\]
That is, $\bar{\rho}_{i}(\mu)$ is the rank of $i$'s least-preferred
outcome among those that are selected by $\mu$ with strictly positive
probability (recall that a higher rank corresponds to a worse school,
and so in our context, the worst-case is given by taking the maximum).
Similarly, $\text{\ensuremath{\underbar{\ensuremath{\rho}}}}_{i}(\mu)$
is the rank of agent $i$'s best outcome over all of the allocations
in the support of $\mu$. 
\begin{defn}
\label{def:NOM}A mechanism $\psi$ is \textbf{not obviously manipulable
(NOM)} if, for any agent $i$ of type $\succ_{i}$ and any $\succ_{i}'\neq\succ_{i}$,
the following are true: 
\begin{enumerate}
\item[(i)] $\max_{\succ_{-i}}\bar{\rho}_{i}(\psi(\succ_{i},\succ_{-i}))\leq\max_{\succ_{-i}}\bar{\rho}_{i}(\psi(\succ_{i}',\succ_{-i}))$ 
\item[(ii)] $\min_{\succ_{-i}}\text{\ensuremath{\underbar{\ensuremath{\rho}}}}_{i}(\psi(\succ_{i},\succ_{-i}))\leq\min_{\succ_{-i}}\text{\ensuremath{\underbar{\ensuremath{\rho}}}}_{i}(\psi(\succ_{i}',\succ_{-i}))$ 
\end{enumerate}
If either of (i) or (ii) does not hold for some agent and type, then
$\succ_{i}'$ is an \textbf{obvious manipulation }for agent $i$ of
type $\succ_{i}$, and the mechanism $\psi$ is said to be \textbf{obviously
manipulable}. 
\end{defn}
To understand the definition of an obvious manipulation, first consider
part (i). On the LHS of the inequality, $\bar{\rho}_{i}(\psi(\succ_{i},\succ_{-i}))$
is the rank of $i$'s worst-case outcome that might arise when the
implemented random allocation is $\psi(\succ_{i},\succ_{-i})$. We
then take the maximum over $\succ_{-i}$. This process returns the
worst-case outcome for $i$ over all possible reports of other agents,
$\succ_{-i}$, and all possible realizations of Nature. The RHS is
the same, just replacing $\succ_{i}$ with a misreport $\succ_{i}'$.
The inequality in (i) says that the rank of the worst-case outcome
under the misreport should be weakly worse (i.e., weakly higher) than
that under the truth. Part (ii) of Definition \ref{def:NOM} is analogous,
except it compares the best-case outcomes instead, and, since lower
numbers correspond to more preferred outcomes, we replace max with
min.

There are several justifications for why a designer might be concerned
with obvious manipulations. As discussed in the introduction, just
because a mechanism \emph{can }be manipulated does not mean that it
\emph{will }be manipulated, and Definition \ref{def:NOM} is a way
to separate those manipulations that are ``obvious'', and are thus
likely to be identified by participants, from those that are not.
Formally, Theorem 1 of \citet{troyan2020obvious} shows that obvious
manipulations are precisely those manipulations that can be recognized
by an agent who is cognitively-limited in the sense defined by \citet{Li-OSP},
and is unable to contingently reason about outcomes state-by-state.
Mathematically, such agents know the range of the function $\psi$
conditional on their own reports, but not the full function itself,
state-by-state.\footnote{This is particularly relevant in the context of school choice. For
instance, \citet{troyan2020obvious} write: ``...this could be a
neighborhood parent group that does not fully understand (or has not
been told) the assignment algorithm but has kept track of what preferences
have been submitted and what the resulting assignments were.'' Such
parent groups are indeed quite common; see \citet{pathak/sonmez:08}.} Allowing some manipulations so long as they are non-obvious widens
the space of mechanisms available to the designer, which may allow
for improvements on other dimensions, such as the average rank. Further,
unlike other relaxations of strategyproofness such as Bayesian incentive
compatibility or approximate notions such as strategyproofness-in-the-large
(SPL, \citealp{azevedo2019strategy}), NOM requires no assumptions
on how preferences are drawn or agent beliefs. Rather, NOM is defined
using best and worst case scenarios, which are likely to be particularly
salient.\footnote{Following \citet{troyan2020obvious}, several other papers have applied
non-obvious manipulability to various settings, including \citet{aziz2021obvious},
\citet{ortega2019obvious}, \citet{archbold2022non}, \citet{troyan2020essentially},
and \citet{cerrone2022school}. Also similar in spirit, though technically
different, is \citet{li2021simple} who show that a designer can sometimes
be better off using a non-SP mechanism even when agents are unsophisticated,
no matter how they resolve their ``strategic confusion''.}

\section{\label{sec:Main-Result}Results}

In this section, I provide my main results on the non-obvious manipulability
of RM mechanisms. An important distinction is whether the RM mechanism
under consideration is a full-support mechanism or not. We start by
considering full-support mechanisms, and then move to discuss other
selection rules. 

\subsection{Full-support RM mechanisms}

Recall that we say a rank-minimizing mechanism has full support if,
for all $\succ_{I}$, $\psi(\succ_{I})$ is a full-support RM allocation,
or, equivalently, $\psi(\succ_{I})(\alpha)>0$ for all $\alpha\in\bar{\mathcal{A}}(\succ_{I})$.
In words, for each preference profile $\succ_{I}$, the mechanism
selects every rank-minimizing allocation with strictly positive probability. 
\begin{thm}
\label{thm:RM-is-NOM}Let $\psi$ be a rank-minimizing mechanism with
full support. Then, $\psi$ is not obviously manipulable. 
\end{thm}
The proof of Theorem \ref{thm:RM-is-NOM} can be found in the appendix.
The full-support assumption requires only that the mechanism place
non-zero probability on each rank-minimizing allocation, but otherwise
the distribution can be arbitrary. One example is the mechanism that,
for each preference profile $\succ_{I}$, selects an allocation uniformly
at random from $\bar{\mathcal{A}}(\succ_{I})$, the set of all rank-minimizing
allocations. We call this mechanism the \emph{uniform rank-minimizing
(URM) mechanism. }

URM is a natural choice from the class of RM mechanisms, and is one
that has received attention in the literature. \citet{Nikzad2022Rank}
shows that URM is Bayesian incentive compatible in markets in which
agent preferences are drawn iid and uniformly at random. \citet{Ortega2022ImprovingEfficiency}
also use the URM implementation of RM in both their theoretical results
comparing RM to TTC and DA, as well as in their simulations.\footnote{An additional issue for implementing URM (or, any full-support RM
mechanism) is computational: formally, the mechanism requires finding
all allocations that minimize the average rank. While it is known
that the problem of finding one RM allocation has complexity $O(n^{3})$
\citep{krokhmal2009random,parviainen2004random}, I am not aware of
any results on the computational complexity of finding\emph{ }all
RM allocations, and am unable to speculate on the impact of this issue
in practice. For instance, \citet{Ortega2022ImprovingEfficiency}
run simulations and counterfactual analysis using Hungarian school
choice data, and use an algorithm that searches for one rank-minimizing
allocation. While this is not exactly equivalent to URM, by perturbing
the algorithm slightly, they are able to find many different rank-minimizing
allocations, and they find that there is little variance across them,
suggesting that this may not be a major issue in practice.} 

By Theorem \ref{thm:RM-is-NOM}, URM is not obviously manipulable.
Besides satisfying appealing efficiency (rank-minimizing) and strategic
(NOM) properties, a final advantage of URM is that because it randomizes
uniformly, it treats all agents fairly. This can be formalized as
follows. Let $\psi_{i}(\succ_{I})$ be agent $i$'s lottery over objects
she is assigned that is induced by the random allocation $\psi(\succ_{I})$.
A mechanism $\psi$ satisfies \textbf{equal treatment of equals (ETE)
}if for all $\succ_{I}\in\mathcal{P}^{I}$ and all $i,j\in I$, $\succ_{i}=\succ_{j}$
implies $\psi_{i}(\succ_{I})=\psi_{j}(\succ_{I})$. 
\begin{prop}
The uniform rank-minimizing mechanism satisfies equal treatment of
equals. 
\end{prop}
The proof of this proposition is simple to see. Consider two agents
$i$ and $j$ that have the same preferences. Because $i$ and $j$
have the same preferences, starting from any allocation and swapping
their assignments results in no change in the average rank. In particular,
for any rank-minimizing allocation $\alpha\in\text{supp}(\psi(\succ_{I}))$,
there is another allocation $\alpha'\in\text{supp}(\psi(\succ_{I}))$
such that $\alpha_{i}'=\alpha_{j}$, $\alpha_{j}'=\alpha_{i}$, and
$\alpha_{k}'=\alpha_{k}$ for all $k\neq i,j$. Since the URM mechanism
selects each element in $\text{supp}(\psi(\succ_{I}))$ with equal
probability, by symmetry, agents $i$ and $j$ will have the same
lottery over final objects. 

\subsection{RM mechanisms without full support}

While full support RM mechanisms---in particular, the URM mechanism---seem
natural, and are always NOM, it is also possible to consider other
RM mechanisms that violate this assumption, and so, it is also necessary
to investigate whether the NOM property extends beyond full-support
mechanisms. As I will show in this section, the answer is ``it depends''. 

I start by showing that full support is not necessary for NOM implementation
by showing that there exist deterministic RM mechanisms that are NOM.\footnote{By deterministic I mean that for each $\succ_{I}$, $\psi(\succ_{I})(\alpha)=1$
for one allocation $\alpha$ and $\psi(\succ_{I})(\alpha')=0$ for
all $\alpha'\neq\alpha.$} In particular, \citet{featherstone2014rank} suggests the following
tie-breaking procedure:\footnote{\citet{featherstone2014rank} applies this procedure to his broader
class of welfare-maximization mechanisms, which contains the class
of RM mechanisms.} find the set of deterministic RM allocations $\bar{\mathcal{A}}(\succ_{I})$,
order the agents in some (exogenous) way, and run a serial dictatorship
starting with $\bar{\mathcal{A}}(\succ_{I})$. In the serial dictatorship
phase, when each agent's turn comes, she finds the remaining allocation(s)
that give her her most preferred object, and eliminates all others.
We call this mechanism the \emph{rank-minimizing serial dictatorship
(RM-SD) mechanism}. It is obvious that the RM-SD mechanism always
ends with a single deterministic allocation at the end of the serial
dictatorship phase, and so this is a deterministic mechanism (in particular,
it does not have full support).

\subsubsection*{Rank-Minimizing Serial Dictatorship (RM-SD)}

For a set of allocations $A'\subseteq\mathcal{A}$ and an agent preference
$\succ_{i}$, let 
\[
Top_{\succ_{i}}(A')=\{\alpha\in A':\alpha_{i}\succsim_{i}\hat{\alpha}_{i}\text{ for all }\hat{\alpha}_{i}\in A'\}.
\]
In words, $Top_{\succ_{i}}(A')$ consists of all of the allocations
in the set $A'$ at which agent $i$ gets her top choice. The RM-SD
mechanism is defined as follows.
\begin{itemize}
\item Fix a bijection $f:\{1,\ldots,N\}\rightarrow I$. This bijection produces
an ordering of the agents, where agent $f(1)$ is first, agent $f(2)$
is second, etc.
\item Given a reported preference profile $\succ_{I}$, let $\bar{\mathcal{A}}(\succ_{I})$
be the set of rank-minimizing allocations at $\succ_{I}$. Initialize
$A_{0}=\bar{\mathcal{A}}(\succ_{I})$. 
\item Consider agent $f(1)$, and calculate the set $A_{1}=Top_{\succ_{f(1)}}(A_{0}).$ 
\item Consider agent $f(2)$, and calculate the set $A_{2}=Top_{\succ_{f(2)}}(A_{1})$
\item etc. 
\end{itemize}
The mechanism ends with a final set $A_{N}=Top_{\succ_{f(N)}}(A_{N-1})$,
where $A_{N}=\{\alpha\}$. That is, $A_{N}$ contains a unique allocation
$\alpha$. This allocation $\alpha$ is the final output of the RM-SD
mechanism.

\medskip

For the next result, I focus on \textbf{unit capacity markets} in
which $|I|=|O|=N$, and $q_{o}=1$ for all $o\in O$.\textbf{ }When
each object has unit capacity, the RM-SD mechanism is not obviously
manipulable.
\begin{prop}
\label{prop:RMSD-isNOM}In unit capacity markets, the RM-SD mechanism
is not obviously manipulable. 
\end{prop}
The proof of this proposition is in the appendix. Proposition \ref{prop:RMSD-isNOM}
shows that the full-support assumption of Theorem \ref{thm:RM-is-NOM}
is not necessary for RM to be NOM. However, the unit capacity assumption
is needed to get this result, and if we do not require it, RM-SD is
no longer NOM.
\begin{prop}
\label{prop:RM-SD-is-obvi-manip}Assume that there are at least 3
objects and at least 4 agents, and at least one object has capacity
$q_{o}>1$. Then, RM-SD is obviously manipulable.
\end{prop}
The proof considers a market with 4 agents and 3 objects with capacities
$q_{1}=q_{3}=1$ and $q_{2}=2$, and focuses on an agent $i$ with
preferences $o_{1}\succ_{i}o_{2}\succ_{i}o_{3}$.\footnote{The same proof can be easily embedded in larger markets. The key feature
is that at least one object must have greater than unit capacity.
Also, the proof of Proposition \ref{prop:RMSD-isNOM} above points
out where the argument no longer holds when moving from unit capacity
to more general markets.} When agent $i$ is second in the serial dictatorship stage, his favorite
object $o_{1}$ might be taken by the first agent, and then the rank-minimizing
constraints relegate him to $o_{3}$, his worst object. If he instead
reports $o_{2}\succ_{1}'o_{1}\succ_{1}'o_{3}$, then, because $o_{2}$
has capacity 2 and agent $i$ is second in the serial dictatorship,
he is able to guarantee himself $o_{2}$, which is an obvious manipulation.
The full details are in the appendix.

I close this section by noting that even in unit capacity markets,
there exist RM mechanisms that are obviously manipulable. 
\begin{prop}
\label{prop:OM-that-is-RM}Even in unit capacity markets, there exist
rank-minimizing mechanisms (without full support) that are obviously
manipulable. 
\end{prop}
The proof of this proposition (also in the appendix) is by again example.
I include this result because the mechanism works differently than
the one used to prove Proposition \ref{prop:RM-SD-is-obvi-manip},
and I think is instructive. In particular, I consider a market of
three agents $I=\{i,j,k\}$ and three objects $O=\{o_{1},o_{2},o_{3}\}$.\footnote{Again, this can be easily embedded in larger markets. }
When all agents report the same preferences, say $o_{1}\succ o_{2}\succ o_{3}$,
all allocations are rank-minimizing. Since we do not require full-support,
the mechanism always selects a single allocation such that agent $i$
receives her worst choice: $\alpha_{i}=o_{3}$. However, when agent
$i$ reports a different preference, say $o_{2}\succ'_{i}o_{1}\succ_{i}'o_{3}$,
the mechanism always selects an allocation in which agent $i$ does
\emph{not }receive her worst choice: $\alpha_{i}'\neq o_{3}$. (The
main work of the proof is to show that no matter the preferences of
the other agents, it is always possible to find a rank-minimizing
allocation where $\alpha_{i}'\neq o_{3}$.) Thus, if agent $i$ reports
her true type $o_{1}\succ_{i}o_{2}\succ_{i}o_{3}$, her worst case
is $o_{3}$, while if she misreports $o_{2}\succ_{i}'o_{1}\succ_{i}'o_{3}$,
her worst case is strictly better than $o_{3}$. Therefore, $\succ_{i}'$
is an obvious manipulation. for type $\succ_{i}$. 

Notice that in this mechanism, agent $i$ is treated differently than
the other agents: effectively, the mechanism ``protects'' $i$ from
ever receiving her worst choice if she reports $\succ_{i}'$. Even
though there may be rank-minimizing allocations at which $i$ receives
$o_{3}$, the mechanism never selects these, whereas a full-support
mechanism sometimes would. Indeed, it seems intuitive that if a mechanism
is protecting an agent from her worst choice at some preference profiles,
but not at others, it will be ``obvious'' that this mechanism can
be manipulated. Because agent $i$ is treated differently than the
others, this mechanism also violates the fairness condition of equal
treatment of equals introduced in the previous subsection, and that
is satisfied by the URM mechanism.\footnote{The RM-SD mechanism also violates equal treatment of equals, as the
agent who goes first in the serial dictatorship phase has an advantage
relative to any later agents who submit the same preferences.}

\medskip

The above results provide just a sampling of possible selection rules
for RM, and there are of course many others. For instance, while RM-SD
is obviously manipulable when objects can have capacity greater than
1, there could be other selection rules for which RM is NOM. While
such a broad characterization is beyond the scope of this paper, the
results here suggest that when going beyond full support RM mechanisms,
the details of the selection rule will matter for the NOM properties
of RM mechanisms.

\section{\label{sec:Discussion-and-Conclusion}Conclusion}

This paper investigates conditions under which RM mechanisms, while
not strategyproof, are at least not obviously manipulable. I show
that as long as the mechanism has full support, RM will be NOM. In
particular, the uniform RM mechanism in which the designer selects
the final allocation uniformly at random from among all of those allocations
that minimize the average rank is an NOM mechanism. While this is
quite a natural selection rule, and one that has been used by other
papers in the literature on RM, there are instances in which a designer
may have some interest in favoring some RM allocations over others,
and thus may find it desirable to use a selection rule that does not
have full-support. My results suggest that care should be taken when
doing so. While such mechanisms may achieve the designer's secondary
objectives while retaining the rank-minimizing property, they might
also compromise the (non-)obvious manipulability of the resulting
mechanism and ultimately undermine these objectives. 

In sum, the rank-minimizing mechanism is appealing for assignment
problems, because it directly optimizes a natural objective that is
desirable to policy-makers. Thus, it is somewhat striking that there
has been thus far relatively little written about this mechanism in
the economics (and in particular, school choice) literature, while
there have been perhaps hundreds of papers written about mechanisms
such as DA and TTC. One possible explanation for this gap is that
the literature is overly-focused on strategyproofness as an incentive
property, which is satisfied by both DA and TTC. While strategyproofness
is a very appealing desideratum, it limits the flexibility for a designer
to optimize on other important dimensions. Considering weaker properties
such as NOM allows access to a broader class of mechanisms, including
(some) rank-minimizing ones. Whether such mechanisms will be manipulated
in practice is ultimately an empirical question, but given my results
and their other desirable properties, I argue that RM mechanisms are
a class of mechanisms that is worthy of further investigation for
practical market design applications.

\section*{Appendix: Proofs}

\subsection*{Proof of Theorem \ref{thm:RM-is-NOM}}

Let $\psi^{RM}$ be a rank-minimizing mechanism with full support.
I start with the following lemma. 
\begin{lem}
\label{lem:same-preferences-characterization}Consider a preference
profile $\succ_{I}'$ such that for $\succ_{i}'=\succ_{j}'$ for all
$i,j\in I$, and wlog, let this preference ranking be $\succ_{j}':o_{1},o_{2},\ldots,o_{M}$.
Define $m^{*}=\min\{m:\sum_{m'=1}^{m}q_{m'}\geq N\}$ and $O'=\{o_{1},\ldots,o_{m^{*}}\}$,
and let $A^{*}\subseteq\mathcal{A}$ be the subset of allocations
that satisfy: 
\begin{enumerate}
\item[(I)] All $o_{m}\in O'\setminus\{o_{m^{*}}\}$ are assigned to exactly
$q_{m}$ agents. 
\item[(II)] Object $o_{m^{*}}$ is assigned to exactly $N-\sum_{m'=1}^{m^{*}-1}q_{m'}$
agents. 
\end{enumerate}
Then, $\bar{\mathcal{A}}(\succ_{I}')=A^{*}$. Further, for any $\tilde{o}\in O'=\{o_{1},\ldots,o_{m^{*}}\}$,
there is at least one allocation $\alpha\in\bar{\mathcal{A}}(\succ_{I}')$
such that $\alpha_{i}=\tilde{o}.$ 
\end{lem}
In words, object $o_{m^{*}}$ is the critical object, in the sense
that the set $O'=\{o_{m_{1}},\ldots,o_{m^{*}}\}$ has enough total
capacity to accommodate all agents, while the set $O'\setminus\{o_{m^{*}}\}$
does not. The set $A^{*}$ is then the set of all allocations that
consist of all possible ways of assigning the $N$ agents to the objects
in $O'$ such that $o_{m_{1}},\ldots,o_{m^{*}-1}$ are filled to capacity,
and all remaining agents are assigned to $o_{m^{*}}$. The lemma says
that when all agents have the same preferences, any rank-minimizing
allocation must satisfy (I) and (II).

\emph{Proof of Lemma \ref{lem:same-preferences-characterization}.}
Under the preference profile given, since all agents rank everything
the same, the average rank of any allocation $\alpha$ is:

\[
\bar{r}(\alpha)=\frac{1}{N}\left(\sum_{m=1}^{M}m|\{i\in I:\alpha_{i}=o_{m}\}|\right).
\]
This is clearly minimized by any allocation that assigns $q_{1}$
students to $o_{1}$, $q_{2}$ students to $o_{2}$, etc., until all
students are assigned. This is precisely the set of allocations $A^{*}$,
and so by construction, any allocation $\alpha\in A^{*}$ achieves
the minimum average rank. For any allocation $\alpha'\notin A^{*}$,
either (I) or (II) must fail. If (I) fails, let $o_{\hat{m}}\in\{o_{1},\ldots,o_{m^{*}-1}\}$
be an object that is not filled to capacity. Then, there must be some
agent $j$ assigned to some object $o_{m^{*}},\ldots,o_{M}$. Reassigning
agent $j$ to object $o_{\hat{m}}$ and leaving all other assignments
the same lowers the average rank, and so $\alpha'\notin\bar{\mathcal{A}}(\succ_{I}).$
Therefore, (I) must hold. If (II) fails, then once again, there is
some agent $j$ assigned to some object $o_{m^{*}+1},\ldots,o_{M}$,
and we can move this agent to object $o_{m^{*}}$ and lower the average
rank. Therefore, $\bar{\mathcal{A}}(\succ_{I}')=A^{*}$. Lastly, because
all agents have the same preferences, it does not matter precisely
which agents are assigned to which objects, and so by symmetry, there
exists at least one $\alpha\in\bar{\mathcal{A}}(\succ_{I}')$ such
that $\alpha_{i}=\tilde{o}$ for all $\tilde{o}\in O'$.

$\hfill\blacksquare$

Now, consider an agent $i$ with type $\succ_{i}$. I show that for
any $\succ_{i}'\neq\succ_{i}$, each part of Definition \ref{def:NOM}
holds for any rank-minimizing mechanism with full support.

\subsubsection*{Part (i): $\max_{\succ_{-i}}\bar{\rho}_{i}(\psi^{RM}(\succ_{i},\succ_{-i}))\protect\leq\max_{\succ_{-i}}\bar{\rho}_{i}(\psi^{RM}(\succ_{i}',\succ_{-i}))$.}

Without loss of generality, index agent $i$'s true type as $\succ_{i}:o_{1},o_{2},\ldots,o_{M}$.
I first show that when $i$ reports her true preferences, $\max_{\succ_{-i}'}\bar{\rho}_{i}(\psi^{RM}(\succ_{i},\succ_{-i}'))=m^{*}$,
where $m^{*}$ is as defined in Lemma \ref{lem:same-preferences-characterization}.
When $\succ_{j}=\succ_{i}$ for all $j$, there is at least one allocation
$\alpha\in\bar{\mathcal{A}}(\succ_{I})$ such that $\alpha_{i}=o_{m^{*}},$
by Lemma \ref{lem:same-preferences-characterization}. Since $\psi^{RM}$
has full support, this implies $\max_{\succ_{-i}'}\bar{\rho}_{i}(\psi^{RM}(\succ_{i},\succ_{-i}'))\geq m^{*}$.
To show equality, assume that $\max_{\succ_{-i}}\bar{\rho}_{i}(\psi^{RM}(\succ_{i},\succ_{-i}))>m^{*}$.
Then, there must be some $\succ_{-i}'$ and some $\alpha'\in\bar{\mathcal{A}}(\succ_{i},\succ_{-i}')$
such that $\alpha_{i}'=o_{m'}$ for some $m'>m^{*}$. This implies
that there is some $m''\leq m^{*}$ such that object $o_{m''}$ is
not filled to capacity. Thus, consider an alternative allocation $\alpha''$
where agent $i$ is reassigned to $o_{m''}$ and all other agents
have the same assignment as in $\alpha'$. Then, we have $\bar{r}(\alpha'')<\bar{r}(\alpha')$,
i.e., this lowers the average rank, which contradicts that $\alpha'\in\bar{\mathcal{A}}(\succ_{i},\succ_{-i}')$.
Therefore, $\max_{\succ_{-i}'}\bar{\rho}_{i}(\psi^{RM}(\succ_{i},\succ_{-i}'))=m^{*}$.

Thus, I have shown that when $i$ reports her true preferences, her
worst-case outcome is $\max_{\succ_{-i}'}\bar{\rho}_{-i}(\psi^{RM}(\succ_{i},\succ_{-i}'))=m^{*}$.
What remains to show is that for any misreport $\succ_{i}'\neq\succ_{i},$
we have $\max_{\succ_{-i}'}\bar{\rho}_{i}(\psi^{RM}(\succ_{i}',\succ_{-i}'))\geq m^{*}$,
where of course the maximum is evaluated with respect to $i$'s true
preferences. Consider a misreport $\succ_{i}'\neq\succ_{i}$. For
notational purposes, index this preference profile as 
\[
\succ_{i}':o_{r_{1}},o_{r_{2}},\ldots,o_{r_{M}}.
\]
Let $m^{**}=\min\{m:\sum_{m'=1}^{m}q_{r_{m'}}\geq N\}$. Similar to
Lemma \ref{lem:same-preferences-characterization}, this is the index
of the critical object in the sense that the set $O''=\{o_{r_{1}},\ldots,o_{r_{m^{**}}}\}$
has enough total capacity for all agents, but the set $\{o_{r_{1}},\ldots,o_{r_{m^{**}-1}}\}$
does not. Consider a preference profile $\succ_{I}'$ such that $\succ_{j}'=\succ_{i}'$
for all $j\in I$. By Lemma \ref{lem:same-preferences-characterization},
we have $\bar{\mathcal{A}}(\succ_{I}')=A{}^{**}$, where $A^{**}$
is defined analogously to $A^{*}$ in Lemma \ref{lem:same-preferences-characterization},
replacing $O'$ with the set $O''$.

\textbf{Case 1: $O''=O'$. }

Since $o_{m^{*}}\in O'=O''$, by Lemma \ref{lem:same-preferences-characterization},
there is at least one allocation $\alpha\in\bar{\mathcal{A}}(\succ_{I}')$
such that $\alpha_{i}=o_{m^{*}}$. Since $\psi^{RM}$ has full support,
$\max_{\succ_{-i}'}\bar{\rho}_{i}(\psi^{RM}(\succ_{i}',\succ_{-i}')\geq m^{*}$,
as desired.

\textbf{Case 2: $O''\neq O'$. }

By definition, we cannot have $O''\subsetneq O'$,\footnote{This follows because $O'=\{o_{1},\ldots,o_{m^{*}}\}$ where $m^{*}$
is the smallest index such that $O'$ contains enough capacity for
all agents. If $O''\subsetneq O'$, then $O''$ cannot contain enough
capacity for all agents, which contradicts the definition of $O''$. } and so, if $O''\neq O'$, there is some $o_{m}\in O''$ such that
$m>m^{*}$, and thus $o_{m^{*}}\succ_{i}o_{m}$ according to $i$'s
true preferences $\succ_{i}$. By Lemma \ref{lem:same-preferences-characterization},
there is at least one allocation $\alpha\in\bar{\mathcal{A}}(\succ_{I}')$
such that $\alpha_{i}=o_{m}$. Because $o_{m^{*}}\succ_{i}o_{m}$
and $\psi^{RM}$ has full support, this implies that $\max_{\succ_{-i}'}\bar{\rho}_{i}(\psi^{RM}(\succ_{i}',\succ_{-i}')>m^{*},$
as desired.

This completes the argument for part (i). \bigskip{}

\subsubsection*{Part (ii): $\min_{\succ_{-i}}\text{\ensuremath{\underbar{\ensuremath{\rho}}}}_{i}(\psi(\succ_{i},\succ_{-i}))\protect\leq\min_{\succ_{-i}}\text{\ensuremath{\underbar{\ensuremath{\rho}}}}_{i}(\psi(\succ_{i}',\succ_{-i}))$.}

Without loss of generality, consider agent $i$ whose preferences
are $\succ_{i}:o_{1},\ldots,o_{M}$. As in part (i), let $m^{*}=\min\{m:\sum_{m'=1}^{m}q_{m}\geq N\}$.
Consider preference profile $\succ_{-i}$ for the other agents constructed
as follows: 
\begin{itemize}
\item Exactly $q_{1}-1$ agents have preferences such that $\succ_{j}:o_{1},\ldots.$ 
\item For all $m'=2,\ldots,m^{*}-1$, exactly $q_{m'}$ agents have preferences
such that $\succ_{j}:o_{m'},\ldots.$ 
\item Exactly $N-\sum_{m'=1}^{m^{*}-1}q_{m'}$ agents have preferences such
that $\succ_{j}:o_{m^{*}},\ldots.$ 
\end{itemize}
In words, the constructed profile $\succ_{I}$ is such that each object
$o_{m}$ has exactly $q_{m}$ agents who have ranked it first. This
is possible by the definition of $m^{*}$ and the assumption that
$\sum_{m}q_{m}\geq N$. Now, notice that at this profile, there is
a unique rank-minimizing allocation, $\psi^{RM}(\succ_{I})=\{\alpha^{*}\}$,
where $\alpha^{*}$ is the allocation such that each agent is assigned
to her first-choice object. Thus, $\text{\ensuremath{\underbar{\ensuremath{\rho}}}}_{i}(\psi(\succ_{i},\succ_{-i}))=1$,
and so $\min_{\succ_{-i}}\text{\ensuremath{\underbar{\ensuremath{\rho}}}}_{i}(\psi(\succ_{i},\succ_{-i}))=1$.
Since it is obvious that $\text{\ensuremath{\underbar{\ensuremath{\rho}}}}_{i}(\psi(\succ_{i}',\succ_{-i}))\geq1$
for any $(\succ_{i}',\succ_{-i})$, we have $\min_{\succ_{-i}}\text{\ensuremath{\underbar{\ensuremath{\rho}}}}_{i}(\psi(\succ_{i},\succ_{-i}))\leq\min_{\succ_{-i}}\text{\ensuremath{\underbar{\ensuremath{\rho}}}}_{i}(\psi(\succ_{i}',\succ_{-i}))$,
and therefore part (ii) of Definition \ref{def:NOM} holds. $\hfill\blacksquare$

\subsection*{Proof of Proposition \ref{prop:RMSD-isNOM}}

Consider a market such that $|I|=|O|=N$, and $q_{o}=1$ for all $o\in O$.
Let $\psi$ denote the RM-SD mechanism with fixed agent ordering $f(1),\ldots,f(N)$
in the SD stage. I start with the following claim. 
\begin{claim}
\label{claim:1-might-get-anything-but-last}Consider an agent, wlog
labeled agent 1, with preferences $o_{1}\succ_{1}o_{2}\succ_{1}\cdots\succ_{1}o_{N}$.\footnote{Agent 1 need not be the first agent in the SD ordering, i.e., $f(1)\neq1$.
Further, the indexing of the true preference ordering as $o_{1}\succ_{1}o_{2}\succ_{1}\cdots$
is without loss of generality. The same arguments will apply to any
agent with any true preference ordering.} For all objects $o\neq o_{N}$, there exists a preference profile
for the remaining agents $\succ_{-1}$ such that there is a unique
rank-minimizing allocation $\alpha$ at $\succ_{I}=(\succ_{1},\succ_{-1})$,
and at this allocation $\alpha_{1}=o$. 
\end{claim}
In other words, this claim says that, for any preferences the agent
submits, under any rank-minimizing mechanism, the agent could receive
any object, with the possible exception of the object she ranks last.

\emph{Proof of Claim }\ref{claim:1-might-get-anything-but-last}.
Wlog, let $o=o_{m}$, where $m\leq N-1$. Consider a profile of preferences
defined as follows:

\[
\begin{array}{c|c|c|c|c|c|c|c|c}
\succ_{1} & \succ_{2} & \succ_{3} & \quad\cdots\quad & \succ_{m} & \succ_{m+1} & \quad\cdots\quad & \succ_{N-1} & \succ_{N}\\
\hline \vdots & \boxed{o_{1}} & \boxed{o_{2}} &  & \boxed{o_{m-1}} & \boxed{o_{m+1}} &  & \boxed{o_{N-1}} & \boxed{o_{N}}\\
\vdots &  &  &  &  &  &  & \\
\boxed{o_{m}} & \vdots & \vdots & \quad\cdots\quad &  &  & \quad\cdots\quad & \vdots & \vdots\\
\vdots &  &  &  &  &  &  & \\
o_{N} & o_{m} & o_{m} &  & o_{m} & o_{m} &  & o_{m} & o_{m}
\end{array}
\]
In other words, all agents besides agent 1 rank $o_{m}$ last, and
each of these $N-1$ agents has a distinct favorite object in the
set $O\setminus\{o_{m}\}$ (the dots indicate that the remaining parts
of the preference profile can be arbitrary). Let the allocation in
boxes be denoted by $\alpha$. 

For any allocation $\alpha'$, let $R(\alpha')=\sum_{i=1}^{N}r_{i}(\alpha_{i}')$
be the total sum of ranks.\footnote{Obviously, minimizing average rank is equivalent to minimizing the
total sum of ranks. I work with the latter here to avoid having to
carry around the $1/N$ notation everywhere. } We claim that $\alpha$ in the boxes is the unique rank-minimizing
allocation for this preference profile. To see this, note that 
\[
R(\alpha)=\overbrace{m}^{\text{Agent 1}}+\overbrace{1+1\cdots+1}^{N-1\text{ times }}=m+(N-1)
\]
It is immediate to see that any alternative allocation $\alpha'$
in which $\alpha'_{1}=o_{m}$ must have $R(\alpha')>R(\alpha)$, because
agent 1's rank remains unchanged, and the total sum of ranks of agents
$2,\ldots,N$ is clearly minimized by giving them all their first
choice. Next, consider an allocation $\alpha'$ such that $\alpha'_{1}=o_{m'}$
for some $m'>m$. At this allocation, $r_{1}(\alpha'_{1})>r_{1}(\alpha_{1})$,
and $r_{j}(\alpha'_{j})\geq1=r_{j}(\alpha_{j})$ for all $j\neq1$,
and thus $R(\alpha')>R(\alpha)$. 

Finally, consider an allocation such that $\alpha_{1}'=o_{m'}$ for
some $m'<m$. We can write:
\[
R(\alpha')-R(\alpha)=\sum_{i=1}^{N}(r_{i}(\alpha'_{i})-r_{i}(\alpha_{i})).
\]
 Notice that $r_{1}(\alpha_{1}')-r_{1}(\alpha_{1})=m'-m$. Since some
agent $j$ must be assigned to $o_{m}$, and all of the other agents
rank it last, we have for the agent $j$ that receives $o_{m}$, $r_{j}(\alpha_{j}')-r_{j}(\alpha_{j})=N-1$.
For all other agents, $r_{k}(\alpha_{k}')-r_{k}(\alpha_{k})\geq0$
(since $r_{k}(\alpha_{k})=1$ and $r_{k}(\alpha_{k}')\geq1$). Thus,
we have 
\[
R(\alpha')-R(\alpha)=m'-m+(N-1)+\sum_{k\neq i,j}(r_{k}(\alpha_{k}')-r_{k}(\alpha_{k}))
\]
The last summation is bounded below by 0, so $R(\alpha')-R(\alpha)\geq m'-m+(N-1)\geq2-N+(N-1)=1$,\footnote{The second inequality follows because $m\leq N-1$ and $m'\geq1$,
which implies that $m'-m\geq2-N$. } i.e., $R(\alpha')>R(\alpha)$. Thus, $\alpha$ is the unique rank-minimizing
allocation, and at this allocation, $\alpha_{1}=o_{m}$. 

$\hfill\blacksquare$

Note that Claim \ref{claim:1-might-get-anything-but-last} does \emph{not
}apply to $o_{N}$ (or, more generally, the object that agent 1 ranks
last). In particular, the preferences used in the proof no longer
work, because when all agents rank the same object last, there might
be multiple rank-minimizing allocations. 

Now, consider an agent 1 with true preferences $o_{1}\succ_{1}o_{2}\succ_{1}\cdots\succ_{1}o_{N}$.
By Claim \ref{claim:1-might-get-anything-but-last}, we have that
\begin{equation}
\max_{\succ_{-1}}\bar{\rho}_{1}(\psi(\succ_{1},\succ_{-1}))\geq N-1.\label{eq:worst-case-truth-atleast-N-1}
\end{equation}
Consider a false report $\succ_{1}'$, and index this report as $p_{1}\succ_{1}'p_{2}\succ_{1}'\cdots\succ_{1}'p_{N}$
(note that $p_{m}$---the $m^{th}$ ranked object for $\succ_{1}'$---may
be different from $o_{m}$, the $m^{th}$ ranked object for the true
preferences $\succ_{1}$). By Claim \ref{claim:1-might-get-anything-but-last}
applied to $\succ_{1}'$, for all $o\neq p_{N}$ there exist preference
profiles $\succ_{-1}'$ such that there is a unique rank-minimizing
allocation $\alpha$ at $(\succ_{1}',\succ_{-1}')$, and at this allocation,
$\alpha_{1}=o$. In particular, this implies that either $o_{N-1}$
or $o_{N}$ is a possible outcome for agent 1, and thus $\max_{\succ_{-1}}\bar{\rho}_{1}(\psi(\succ_{1}',\succ_{-1}))\geq N-1$.
If equation \ref{eq:worst-case-truth-atleast-N-1} holds with equality,
then the proof is complete, as $\max_{\succ_{-1}}\bar{\rho}_{1}(\psi(\succ_{1}',\succ_{-1}))\geq N-1=\max_{\succ_{-1}}\bar{\rho}_{1}(\psi(\succ_{1},\succ_{-1}))$. 

Thus, we need to last consider the case that equation \eqref{eq:worst-case-truth-atleast-N-1}
is a strict inequality, which means that $\max_{\succ_{-1}}\bar{\rho}_{1}(\psi(\succ_{1},\succ_{-1}))=N$.
In this case, when agent 1 submits her true preferences $o_{1}\succ_{1}\cdots\succ_{1}o_{N}$,
there is some preference profile $\succ_{-1}$ where she might receive
object $o_{N}$. We must show that, for any $\succ_{1}'$ that agent
1 might submit, there exists some $\succ_{-1}'$ where she receives
object $o_{N}$. If $\succ_{1}'$ ranks $o_{N}$ anything other than
last, then, as in the previous paragraph, we can apply Claim \ref{claim:1-might-get-anything-but-last}
to the preference $\succ_{1}'$ and conclude there exists some $\succ_{-1}'$
such that agent 1 receives $o_{N}$ for sure, and we are done.

Thus, consider the case that $\succ_{1}'$ ranks $o_{N}$ last. Let
$\succ_{-1}$ be the preference profile of the remaining agents such
that, when 1 reports the truth, agent 1 receives $o_{N}$. In the
RM-SD mechanism at $(\succ_{1}',\succ_{-1})$, when it is agent $1$'s
turn to choose in the serial dictatorship step, it must be that all
remaining allocations she can select from assign her to $o_{N}$ (as
otherwise, she would eliminate these allocations and choose a better
object for herself). In particular, this implies that $f(1)>1$, i.e.,
agent 1 cannot select first.\footnote{\label{fn:f(1)=00005Cneq1}Indeed, if $f(1)=1$, then it must be that
agent 1 receives $o_{N}$ at all allocations in $\bar{\mathcal{A}}(\succ_{I}')$,
and thus some other agent $j$ receives $o_{1}$ at some allocation
$\alpha\in\bar{\mathcal{A}}(\succ_{I}')$. Let $\alpha'$ denote the
assignment at which agent 1 and agent $j$ swap, and all other agents'
assignments remain the same. Then, $\alpha'$ cannot have a worse
average rank than $\alpha$, and so $\alpha'\in\bar{\mathcal{A}}(\succ_{I}')$
as well, and so agent 1 would eliminate $\alpha$ if she chose first. } 

Index agent 1's false report $\succ_{1}'$ as:
\[
\succ_{1}':p_{1},\ldots,p_{N}.
\]

Notice that $p_{n}$ need not be equal to $o_{n}$; we use $p$'s
to convey that this is a report that ranks the objects differently
than the true report (which was indexed $\succ_{1}:o_{1},\ldots,o_{N}$)
while still being able to make easy reference to the $n^{th}$ ranked
object. However, we are in the case that $\succ_{1}'$ ranks $o_{N}$---agent
1's worst object according to her true preferences---last, so $p_{N}=o_{N}$. 

Let $k=f(1)$ the order in which agent 1 chooses in the SD phase.
By footnote \ref{fn:f(1)=00005Cneq1}, $k>1$ and, without loss of
generality, assume that the (exogenous) SD ordering of the agents
is $2,3,\ldots,k,1,k+1,\ldots,N$ (in other words, the ordering not
including agent 1 is just $2,3,4,\ldots,$ and agent 1 is slotted
in the $k^{th}$ position). Consider a preference profile $\succ_{I}'=(\succ_{1}',\succ_{-1}')$
that takes the form shown in Table \ref{tab:Preference-profile-used}.

\begin{table}
\[
\begin{array}{c|c|c|c|c}
\succ_{1}' & \succ_{2}' & \succ_{3}' & \quad\cdots\quad & \succ_{N}'\\
\hline p_{1} & \boxed{p_{1}} & \boxed{p_{2}} &  & \boxed{p_{N-1}}\\
\vdots & \vdots & \vdots &  & \vdots\\
\boxed{p_{N}} & p_{N} & p_{N} &  & p_{N}
\end{array}
\]

\caption{Preference profile used for proving part of Proposition \ref{prop:RMSD-isNOM}.
Recall that $p_{N}=o_{N}$.\label{tab:Preference-profile-used} The
boxed allocation represents the outcome of the RM-SD mechanism at
this preference profile. }
\end{table}

\begin{claim}
For any rank-minimizing allocation $\alpha\in\bar{\mathcal{A}}(\succ_{I}')$,
object $p_{N}$ is assigned to either agent 1 or agent 2. 
\end{claim}
\emph{Proof.} First, notice that every agent ranks $p_{N}$ last,
and it must be given to someone. Thus, the lowest possible total sum
of ranks, $N+(N-1)\times1=2N-1$, and indeed, this is achievable by,
for instance, the allocation in boxes in the table. What we show is
that at any allocation $\alpha'$ such that $\alpha_{j}=p_{N}$ for
some $j\neq1,2$, $R(\alpha')>2N-1$. Write 
\begin{align*}
R(\alpha') & =(r'_{1}(\alpha_{1}')+r_{2}'(\alpha_{2}'))+(r_{j}'(\alpha_{j}'))+\left(\sum_{i\neq1,2,j}r_{i}'(\alpha_{i}')\right)\\
 & \geq(3)+(N)+(N-3)\\
 & =2N\\
 & >2N-1.
\end{align*}
 The first inequality follows because (i) only one of agent 1 or 2
can be getting their first choice,\footnote{This is the point at which the proof breaks down for the general case
(beyond unit capacity). In the general case, it is possible that the
first choice of agent 1 and 2 has more than one unit of capacity,
and so they both might receive it.} and so $r'_{1}(\alpha_{1}')+r_{2}'(\alpha_{2}')\geq3$; (ii) by assumption,
$r'_{j}(\alpha'_{j})=N$ and (iii) $\sum_{i\neq1,2,j}r_{i}'(\alpha_{i}')\geq N-3$.
$\hfill\blacksquare$

Now, agent 2 chooses ahead of agent 1 in the serial dictatorship phase
of the mechanism, and so will eliminate all allocations in $\bar{\mathcal{A}}(\succ_{I}')$
that give her $p_{N}$. Thus, when it is agent 1's turn to choose,
all remaining allocations are such that $\alpha_{1}=p_{N}$. Recalling
that $p_{N}=o_{N}$, we have shown that for any false report $\succ_{1}':p_{1},\ldots,p_{N}$,
there is a preference profile of the other agents $\succ_{-1}'$ such
that at $(\succ_{1}',\succ_{-1}')$ agent 1 gets $o_{N}$. 

We have therefore shown that, for all $\succ_{1}'$, $\max_{\succ_{-1}}\bar{\rho}_{1}(\psi(\succ_{1}',\succ_{-1}))\geq\max_{\succ_{-1}}\bar{\rho}_{1}(\psi(\succ_{1},\succ_{-1}))$,
which is part (i) of the definition of NOM. Part (ii) follows trivially:
when all agents have a unique first choice, the unique rank-minimizing
allocation assigns them each their first choice. So, for any true
preferences $\succ_{1}:o,o',o'',\cdots$, $\min_{\succ_{-1}}\underline{\rho}_{1}(\psi(\succ_{1},\succ_{-1}))=1$,
and it is immediate that $\min_{\succ_{-1}}\underline{\rho}_{1}(\psi(\succ_{1},\succ_{-1}))\leq\min\underline{\rho}_{1}(\psi(\succ_{1}',\succ_{-1}))$.
$\hfill\blacksquare$

\subsection*{Proof of Proposition \ref{prop:RM-SD-is-obvi-manip}}

Let $I=\{i,j,k,\ell\}$, $O=\{o_{1},o_{2},o_{3}\}$ and $q_{1}=q_{3}=1$,
while $q_{2}=2$. Consider a serial dictatorship ordering that is
$j,i,k,\ell$. Let agent $i$'s true preferences be $\succ_{i}:o_{1},o_{2},o_{3}$,
and consider a preference profile given in the following table: 

\[
\begin{array}{c|c|c|c}
\succ_{i} & \succ_{j} & \succ_{k} & \succ_{\ell}\\
\hline o_{1} & \boxed{o_{1}} & \boxed{o_{2}} & \boxed{o_{2}}\\
o_{2} & o_{2} & o_{1} & o_{1}\\
\boxed{o_{3}} & o_{3} & o_{3} & o_{3}
\end{array}
\]
 At these preferences, the boxed allocation, denoted $\alpha$, is
easily seen to be rank-minimizing. Since $j$ chooses first in the
SD phase of the mechanism, she will eliminate all $\alpha'\in\bar{\mathcal{A}}(\succ_{I})$
such that $\alpha'_{j}\neq o_{1}$. 
\begin{claim}
Let $\alpha'\in\bar{\mathcal{A}}(\succ_{I})$ be such that $\alpha_{j}'=o_{1}$.
Then, $\alpha_{i}'=o_{3}$. 
\end{claim}
To see this claim, simply note that if $\alpha_{i}'=o_{2}$, then
one of $k$ or $\ell$ must be assigned $o_{3}$, and so the total
sum of ranks at $\alpha'$ is 7. The allocation $\alpha$ denoted
in boxes has total sum of ranks equal to 6, and so $\alpha'$ is not
rank-minimizing, which is a contradiction. 

The upshot of the above claim is that after $j$ moves in the serial
dictatorship phase of the mechanism, the only allocations remaining
for $i$ to choose from at the second step have $\alpha'_{i}=o_{3}$.
Thus, we have $\max_{\succ_{-i}}\bar{\rho}(\psi(\succ_{i},\succ_{-i}))=3$. 

Consider a false report $\succ_{i}':o_{2},o_{1},o_{3}$. We claim
that $\max_{\succ_{-i}}\bar{\rho}(\psi(\succ_{i}',\succ_{-i}))<3$.
Assume not, i.e., $\max_{\succ_{-i}}\bar{\rho}(\psi(\succ_{i}',\succ_{-i}))=3$,
and choose some $\succ_{-i}'$ such that $\bar{\rho}(\psi(\succ_{i}',\succ_{-i}'))=3.$
Let the set of available allocations when it is $i$'s turn to choose
in the serial dictatorship be $A'$. Since $\bar{\rho}(\psi(\succ_{i}',\succ_{-i}'))=3$,
this implies that for all $\alpha'\in A'$, we have $\alpha_{i}=o_{3}$,
which implies that both $o_{1}$ and $o_{2}$ are filled to capacity
with other agents. In particular, at all $\alpha'\in A'$, there are
two agents assigned to $o_{2}$, and, since $i$ chooses 2nd in the
serial dictatorship, at least one of these agents must be ordered
after $i$. Assume this agent is $k$, i.e., $\alpha'_{k}=o_{2}$
(the same argument works for agent $\ell$). Now, agent $k$ must
rank $o_{2}\succ_{k}o_{3}$, which means that the preference profile
must be one of the following:

\[
\begin{array}{c|c}
\succ_{i}' & \succ_{k}'\\
\hline *o_{2} & o_{1}\\
o_{1} & \boxed{o_{2}}\\
\boxed{o_{3}} & *o_{3}
\end{array}\quad\quad\begin{array}{c|c}
\succ_{i}^{'} & \succ_{k}'\\
\hline *o_{2} & \boxed{o_{2}}\\
o_{1} & *o_{3}\\
\boxed{o_{3}} & o_{1}
\end{array}\quad\quad\begin{array}{c|c}
\succ_{i}^{'} & \succ_{k}'\\
\hline *o_{2} & \boxed{o_{2}}\\
o_{1} & o_{1}\\
\boxed{o_{3}} & *o_{3}
\end{array}.
\]
 In the tables, the boxes denote the assignment $\alpha'$, while
the stars denote an alternative assignment $\alpha^{*}$ where $i$
and $j$ swap their objects, and everything else remains unchanged
(we do not show the assignments of the other agents in the tables).
Notice that in the left two panels, swapping the assignments of $i$
and $j$ strictly lowers the average rank, which means that in fact,
the preferences must be that in the right panel. 

To summarize: the preferences of $i$ and $k$ are in the following
table, and the assignment $\alpha'$ in boxes is such that $\alpha'\in A'$:

\[
\begin{array}{c|c}
\succ_{i}^{'} & \succ_{k}'\\
\hline *o_{2} & \boxed{o_{2}}\\
o_{1} & o_{1}\\
\boxed{o_{3}} & *o_{3}
\end{array}
\]
 Since $\alpha'\in A'$, this means that $\alpha'$ is rank-minimizing
at $\succ_{I}'$. As $\alpha^{*}$ has the same average rank as $\alpha'$,
$\alpha^{*}$ is also rank-minimizing at $\succ_{I}'$. Further, since
the assignments of all other agents remain the same, $\alpha^{*}$
is not eliminated by agent $j$ at the first step of the serial dictatorship.
Thus, $\alpha^{*}$ is in $i$'s opportunity set when it is her turn
to choose, and thus she would choose it and receive $o_{2}$, which
contradicts that $\max_{\succ_{-i}}\bar{\rho}(\psi(\succ_{i}',\succ_{-i}))=3$.
$\hfill\blacksquare$ 

\subsection*{Proof of Proposition \ref{prop:OM-that-is-RM}}

\textbf{Proof}. The proof is by example in a market with three students
$I=\{i,j,k\}$ and three objects $O=\{o_{1},o_{2},o_{3}\}$. We will
build a RM mechanism $\psi$ that is obviously manipulable. To start,
consider the following profile of preferences, $\succ_{I}^{1}$: 

\[
\begin{array}{c|c|c}
\succ_{i}^{1} & \succ_{j}^{1} & \succ_{k}^{1}\\
\hline o_{1} & o_{1} & \boxed{o_{1}}\\
o_{2} & \boxed{o_{2}} & o_{2}\\
\boxed{o_{3}} & o_{3} & o_{3}
\end{array}
\]
Denote the allocation in the boxes by $\alpha$. Notice that all agents
have the exact same preferences, so any allocation minimizes the average
rank, and a rank-minimizing mechanism can select any allocation at
this preference profile. In particular, we set $\psi(\succ_{I}^{1})(\alpha)=1$,
and $\psi(\succ_{I}^{1})(\alpha')=0$ for all $\alpha'\neq\alpha.$ 

Next, consider the following preferences for agent $i$, $\succ_{i}^{2}$:

\[
\begin{array}{c}
\succ_{i}^{2}\\
\hline o_{2}\\
o_{1}\\
o_{3}
\end{array}
\]

\begin{claim}
\label{claim:i-doesn't-get-o3} For all $\succ_{-i}'$, there exists
an $\alpha'$ such that (i) $\alpha'$ is rank-minimizing with respect
to $(\succ_{i}^{2},\succ_{-i}')$ and (ii) $\alpha_{i}'\neq o_{3}$.
\end{claim}
\emph{Proof of Claim \ref{claim:i-doesn't-get-o3}. }Assume not, i.e.,
there exists a preference profile $\succ_{-i}'$ such that all rank-minimizing
allocations with respect to $(\succ_{i}^{2},\succ_{-i}')$ assign
$i$ to $o_{3}$. Let $\alpha'$ be one such rank-minimizing allocation,
and thus $\alpha_{i}'=o_{3}$. Under $\alpha'$, either agent $j$
or $k$ must be assigned to $o_{2}$; wlog, assume that $\alpha_{j}'=o_{2}$.
Notice that $j$ must rank $o_{2}\succ_{j}o_{3}$; if not, then $\alpha'$
is not Pareto efficient ($i$ and $j$ can engage in a Pareto-improving
trade), and so is also not rank-minimizing. Thus, the overall preference
profile must be one of the following (where $k$ receives $o_{1}$
by construction, but $k$'s actual preferences do not matter for the
argument, and so are indicated by dots):

\[
\begin{array}{c|c|c}
\succ_{i}^{2} & \succ_{j}' & \succ_{k}'\\
\hline *o_{2} & o_{1} & \vdots\\
o_{1} & \boxed{o_{2}}\\
\boxed{o_{3}} & *o_{3}
\end{array}\quad\quad\begin{array}{c|c|c}
\succ_{i}^{2} & \succ_{j}' & \succ_{k}'\\
\hline *o_{2} & \boxed{o_{2}} & \vdots\\
o_{1} & *o_{3}\\
\boxed{o_{3}} & o_{1}
\end{array}\quad\quad\begin{array}{c|c|c}
\succ_{i}^{2} & \succ_{j}' & \succ_{k}'\\
\hline *o_{2} & \boxed{o_{2}} & \vdots\\
o_{1} & o_{1}\\
\boxed{o_{3}} & *o_{3}
\end{array}
\]
 The boxes in each table indicate the allocation $\alpha'$, and the
stars indicate the alternative allocation where $i$ and $j$ swap:
$\alpha_{i}^{*}=o_{2},$ $\alpha_{j}^{*}=o_{3},$ and $\alpha_{k}^{*}=\alpha_{k}'=o_{1}$.
Notice that in the two leftmost panels, $\bar{r}(\alpha^{*})<\bar{r}(\alpha')$,
which contradicts that $\alpha'$ was rank-minimizing. In the rightmost
panel, $\bar{r}(\alpha^{*})=\bar{r}(\alpha')$. Since $\alpha'$ was
assumed to be rank-minimizing, $\alpha^{*}$ is also rank-minimizing.
However, this contradicts that all rank-minimizing allocations assign
$i$ to $o_{3}$, and completes the proof of the claim. 

$\hfill\blacksquare$

The upshot of Claim \ref{claim:i-doesn't-get-o3} is that we can construct
an RM mechanism such that, for all $\succ_{-i}'$ and all $\alpha'$
such that $\alpha_{i}'=o_{3}$, we have $\psi(\succ_{i}^{2},\succ_{-i}')(\alpha')=0$.\footnote{This follows because, by the claim, there is always at least one alternative
rank-minimizing assignment where $i$ does not receive $o_{3}$, and
so we can build the mechanism to only put positive probability on
such assignments.} So, consider agent $i$ of type $\succ_{i}^{1}$. If she reports
her true type, she receives $o_{3}$ with probability 1, while if
she reports $\succ_{i}^{2}$, she receives $o_{3}$ with probability
0, and thus, receives some strictly preferred object with probability
1. Since the worst case from reporting $\succ_{i}^{2}$ is strictly
preferred to $o_{3}$, while the worst case from reporting $\succ_{i}^{1}$
(the truth) is $o_{3}$, this is an obvious manipulation, and mechanism
$\psi$ is obviously manipulable.

$\hfill\blacksquare$


\begin{thebibliography}{27}
\newcommand{\enquote}[1]{``#1''}
\expandafter\ifx\csname natexlab\endcsname\relax\def\natexlab#1{#1}\fi

\bibitem[\protect\citeauthoryear{Abdulkadiro{\v g}lu, Che, Pathak, Roth, and
  Tercieux}{Abdulkadiro{\v g}lu et~al.}{2020}]{Abdulkadiro_lu_2020}
\textsc{Abdulkadiro{\v g}lu, A., Y.-K. Che, P.~A. Pathak, A.~E. Roth, and
  O.~Tercieux} (2020): \enquote{Efficiency, Justified Envy, and Incentives in
  Priority-Based Matching,} \emph{American Economic Review: Insights}, 2,
  425--442.

\bibitem[\protect\citeauthoryear{Abdulkadiro\u{g}lu and
  S\"{o}nmez}{Abdulkadiro\u{g}lu and S\"{o}nmez}{2003}]{abdul2003school}
\textsc{Abdulkadiro\u{g}lu, A. and T.~S\"{o}nmez} (2003): \enquote{School
  Choice: A Mechanism Design Approach,} \emph{American Economic Review}, 93,
  729--747.

\bibitem[\protect\citeauthoryear{Archbold, de~Keijzer, and Ventre}{Archbold
  et~al.}{2022}]{archbold2022non}
\textsc{Archbold, T., B.~de~Keijzer, and C.~Ventre} (2022):
  \enquote{Non-Obvious Manipulability for Single-Parameter Agents and Bilateral
  Trade,} \emph{arXiv preprint arXiv:2202.06660}.

\bibitem[\protect\citeauthoryear{Azevedo and Budish}{Azevedo and
  Budish}{2019}]{azevedo2019strategy}
\textsc{Azevedo, E.~M. and E.~Budish} (2019): \enquote{Strategy-proofness in
  the large,} \emph{The Review of Economic Studies}, 86, 81--116.

\bibitem[\protect\citeauthoryear{Aziz and Lam}{Aziz and
  Lam}{2021}]{aziz2021obvious}
\textsc{Aziz, H. and A.~Lam} (2021): \enquote{Obvious Manipulability of Voting
  Rules,} in \emph{International Conference on Algorithmic Decision Theory},
  Springer, 179--193.

\bibitem[\protect\citeauthoryear{Budish and Cantillon}{Budish and
  Cantillon}{2012}]{budish/cantillon:10}
\textsc{Budish, E. and E.~Cantillon} (2012): \enquote{The Multi-unit Assignment
  Problem: Theory and Evidence from Course Allocation at Harvard,}
  \emph{American Economic Review}, 102, 2237--71.

\bibitem[\protect\citeauthoryear{Budish and Kessler}{Budish and
  Kessler}{2017}]{budish2014changing}
\textsc{Budish, E.~B. and J.~B. Kessler} (2017): \enquote{Can Agents ``Report
  Their Types''? An Experiment that Changed the Course Allocation Mechanism at
  Wharton,} \emph{Chicago Booth Research Paper}.

\bibitem[\protect\citeauthoryear{Cerrone, Hermstr{\"u}wer, and Kesten}{Cerrone
  et~al.}{2023}]{cerrone2022school}
\textsc{Cerrone, C., Y.~Hermstr{\"u}wer, and O.~Kesten} (2023): \enquote{School
  choice with consent: an experiment,} \emph{Economic Journal}, forthcoming.

\bibitem[\protect\citeauthoryear{Combe, Tercieux, and Terrier}{Combe
  et~al.}{2022}]{combe2018design}
\textsc{Combe, J., O.~Tercieux, and C.~Terrier} (2022): \enquote{The design of
  teacher assignment: Theory and evidence,} \emph{Review of Economic Studies,
  forthcoming}.

\bibitem[\protect\citeauthoryear{Delacr{\'e}taz, Kominers, and
  Teytelboym}{Delacr{\'e}taz et~al.}{2023}]{delacretaz2016refugee}
\textsc{Delacr{\'e}taz, D., S.~D. Kominers, and A.~Teytelboym} (2023):
  \enquote{Matching mechanisms for refugee resettlement,} \emph{American
  Economic Review}, 113, 2689--2717.

\bibitem[\protect\citeauthoryear{Demeulemeester and Pereyra}{Demeulemeester and
  Pereyra}{2022}]{demeulemeester2022rawlsian}
\textsc{Demeulemeester, T. and J.~S. Pereyra} (2022): \enquote{Rawlsian
  Assignments,} \emph{arXiv preprint arXiv:2207.02930}.

\bibitem[\protect\citeauthoryear{Featherstone}{Featherstone}{2020}]{featherstone2014rank}
\textsc{Featherstone, C.~R.} (2020): \enquote{Rank efficiency: Investigating a
  widespread ordinal welfare criterion,} Working paper.

\bibitem[\protect\citeauthoryear{Gale and Shapley}{Gale and
  Shapley}{1962}]{Gale:AMM:1962}
\textsc{Gale, D. and L.~S. Shapley} (1962): \enquote{College Admissions and the
  Stability of Marriage,} \emph{The American Mathematical Monthly}, 69, 9--15.

\bibitem[\protect\citeauthoryear{Kesten}{Kesten}{2010}]{kesten:QJE:2010}
\textsc{Kesten, O.} (2010): \enquote{School Choice with Consent,}
  \emph{Quarterly Journal of Economics}, 125, 1297--1348.

\bibitem[\protect\citeauthoryear{Krokhmal and Pardalos}{Krokhmal and
  Pardalos}{2009}]{krokhmal2009random}
\textsc{Krokhmal, P.~A. and P.~M. Pardalos} (2009): \enquote{Random assignment
  problems,} \emph{European Journal of Operational Research}, 194, 1--17.

\bibitem[\protect\citeauthoryear{Li and Dworczak}{Li and
  Dworczak}{2021}]{li2021simple}
\textsc{Li, J. and P.~Dworczak} (2021): \enquote{Are simple mechanisms optimal
  when agents are unsophisticated?} in \emph{Proceedings of the 22nd ACM
  Conference on Economics and Computation}, 685--686.

\bibitem[\protect\citeauthoryear{Li}{Li}{2017}]{Li-OSP}
\textsc{Li, S.} (2017): \enquote{Obviously Strategy-Proof Mechanisms,}
  \emph{American Economic Review}, 107, 3257--87.

\bibitem[\protect\citeauthoryear{Nikzad}{Nikzad}{2022}]{Nikzad2022Rank}
\textsc{Nikzad, A.} (2022): \enquote{Rank-optimal assignments in uniform
  markets,} \emph{Theoretical Economics}, 17, 25--55.

\bibitem[\protect\citeauthoryear{Ortega and Klein}{Ortega and
  Klein}{2023}]{Ortega2022ImprovingEfficiency}
\textsc{Ortega, J. and T.~Klein} (2023): \enquote{The cost of
  strategy-proofness in school choice,} \emph{Games and Economic Behavior},
  141, 515--528.

\bibitem[\protect\citeauthoryear{Ortega and Segal-Halevi}{Ortega and
  Segal-Halevi}{2022}]{ortega2019obvious}
\textsc{Ortega, J. and E.~Segal-Halevi} (2022): \enquote{Obvious manipulations
  in cake-cutting,} \emph{Social Choice and Welfare}.

\bibitem[\protect\citeauthoryear{Parviainen}{Parviainen}{2004}]{parviainen2004random}
\textsc{Parviainen, R.} (2004): \enquote{Random assignment with integer costs,}
  \emph{Combinatorics, Probability and Computing}, 13, 103--113.

\bibitem[\protect\citeauthoryear{Pathak and S\"{o}nmez}{Pathak and
  S\"{o}nmez}{2008}]{pathak/sonmez:08}
\textsc{Pathak, P.~A. and T.~S\"{o}nmez} (2008): \enquote{Leveling the Playing
  Field: Sincere and Sophisticated Players in the Boston Mechanism,} 98,
  1636--1652.

\bibitem[\protect\citeauthoryear{Roth and Peranson}{Roth and
  Peranson}{1999}]{roth/peranson:99}
\textsc{Roth, A.~E. and E.~Peranson} (1999): \enquote{The Redesign of the
  Matching Market for American Physicians: Some Engineering Aspects of Economic
  Design,} 89, 748--780.

\bibitem[\protect\citeauthoryear{Sethuraman}{Sethuraman}{2022}]{sethuraman2022Rank}
\textsc{Sethuraman, J.} (2022): \enquote{A note on the average rank of
  rank-optimal assignments,} Working paper, Columbia University.

\bibitem[\protect\citeauthoryear{Shapley and Scarf}{Shapley and
  Scarf}{1974}]{shapley/scarf:74}
\textsc{Shapley, L. and H.~Scarf} (1974): \enquote{On Cores and
  Indivisibility,} \emph{Journal of Mathematical Economics}, 1, 23--37.

\bibitem[\protect\citeauthoryear{Troyan, Delacr{\'e}taz, and
  Kloosterman}{Troyan et~al.}{2020}]{troyan2020essentially}
\textsc{Troyan, P., D.~Delacr{\'e}taz, and A.~Kloosterman} (2020):
  \enquote{Essentially stable matchings,} \emph{Games and Economic Behavior},
  120, 370--390.

\bibitem[\protect\citeauthoryear{Troyan and Morrill}{Troyan and
  Morrill}{2020}]{troyan2020obvious}
\textsc{Troyan, P. and T.~Morrill} (2020): \enquote{Obvious manipulations,}
  \emph{Journal of Economic Theory}, 185, 104970.

\end{thebibliography}
\end{document}